\documentclass[fleqn,usenatbib]{mnras}

\usepackage[T1]{fontenc}

\DeclareRobustCommand{\VAN}[3]{#2}
\let\VANthebibliography\thebibliography
\def\thebibliography{\DeclareRobustCommand{\VAN}[3]{##3}\VANthebibliography}

\usepackage{graphicx}	
\usepackage{amsmath}	
\usepackage{amssymb}	
\usepackage{subfig}
\usepackage{newtxtext,newtxmath}

\newcommand{\thesource}{4XMM J235747.9-323457}

\title[A new candidate pulsating ULX in NGC 7793]{A new candidate pulsating ULX in NGC 7793}

\author[E. Quintin, N. A. Webb, ...]{
E. Quintin,$^{1}$\thanks{E-mail: erwan.quintin@irap.omp.eu}
N. A. Webb,$^{1}$
A. G\'urpide,$^{1}$
M. Bachetti$^{2}$
and F. F\"urst$^{3}$
\\
$^{1}$IRAP, Université de Toulouse, CNRS, CNES, 9 avenue du Colonel Roche, 31028 Toulouse, France\\
$^{2}$INAF-Osservatorio Astronomico di Cagliari, via della Scienza 5, I-09047 Selargius (CA), Italy\\
$^{3}$Quasar Science Resources SL for ESA, European Space Astronomy Centre (ESAC), Science Operations Departement, 28692 Villanueva de la Ca\~nada, Madrid, Spain
}

\date{Accepted XXX. Received YYY; in original form ZZZ}

\pubyear{2020}

\begin{document}
\label{firstpage}
\pagerange{\pageref{firstpage}--\pageref{lastpage}}
\maketitle

\begin{abstract}
We report here the discovery of NGC 7793 ULX-4, a new transient ultraluminous X-ray source (ULX) in NGC 7793, a spiral galaxy already well known for harbouring several ULXs. This new source underwent an outburst in 2012, when it was detected by \textit{XMM-Newton} and the \textit{Swift} X-ray telescope. The outburst reached a peak luminosity of 3.4$\times 10^{39}$ erg\ s$^{-1}$ and lasted for about 8 months, after which the source went below a luminosity of $10^{37}$ erg\ s$^{-1}$; previous \textit{Chandra} observations constrain the low-state luminosity below $\sim$ 2$\times 10^{36}$ erg\ s$^{-1}$, implying a variability of at least a factor 1000. We propose four possible  optical counterparts, found in archival HST observations of the galaxy. A pulsation in the \textit{XMM-Newton} signal was found at 2.52 Hz, with a significance of $\sim3.4\,\sigma$, and an associated spin-up of $\dot{f} = 3.5\times10^{-8}$ Hz.s$^{-1}$. NGC 7793 is therefore the first galaxy to host more than one pulsating ULX.
\end{abstract}




\section{Introduction}
Ultraluminous X-ray sources (ULXs) are extragalactic non-nuclear sources whose luminosity exceed the Eddington limit for a stellar mass compact object ($\sim3\times10^{39}$ erg.s$^{-1}$), while their non-nuclear position excludes AGNs. They were first discovered using the \textit{Einstein Laboratory} \citep{Long1981}, the term being coined later by \citet{Makishima2000}. To explain such high luminosities without exceeding the Eddington limit, these sources were at first thought to reveal the presence of particularly massive black holes \citep[e.g.][]{Colbert1999}. However, two discoveries in the following years challenged this interpretation of ultraluminous sources. The first \textit{NuSTAR} observations of ULXs in the hard X-rays allowed to detect a high-energy cutoff, which was thought to indicate super-Eddington accretion \citep{Bachetti2013,Walton2013}. Then, the discovery of coherent pulsations in a ULX in M82 \citep{Bachetti2014} lead to the conclusion that at least some ULXs could host accreting neutron stars, which are much less massive. This implies that the accretion rate in these sources goes well above the Eddington limit, up to a factor 500 for the pulsating ULX in NGC 5907, for instance \citep{Israel20175907}. Thus, ULXs are precious tools for understanding accretion mechanisms beyond simple disk models \citep{Shakura1973}. For a thorough review of the link between ULXs and Super Eddington accretion, see for example \citet{Kaaret2017}.

To date, only 7 pulsating ULXs have been found: M82 X-2 \citep{Bachetti2014}, NGC7793 P13 \citep{Furst2016}, NGC5907 ULX-1 \citep{Israel20175907}, NGC300 ULX-1 \citep{Carpano2018}, NGC1313 X-2 \citep{Sathyaprakash2019}, M51 ULX7 \citep{rodriguezcastilloDiscoveryPulsarDay2020}. Swift J0243.6+6124 has been proposed as the first pulsating ULX in our own galaxy \citep{Wilson-Hodge2018}. These sources have several common properties: mainly, their extreme luminosities (even among ULXs), and their long-term temporal variability. Indeed, they exhibit two levels of emission, with drops of up to three orders of magnitude in flux between high-activity and low-activity episodes \citep[see for instance][]{Walton2015}. This common feature lead to the use of the long term variability of ULXs as a proxy to find more pulsating ULXs \citep{Earnshaw2018,Song2020}. Identifying the ULXs with neutron star accretors would allow us to determine the fraction of ULXs that are powered by accretion onto neutron stars.

NGC 7793 P13 is one of those pulsating ULXs. It was discovered during the first X-ray survey of NGC 7793 by \citet{Read1999}, using the ROSAT telescope. This spiral galaxy lies in the Sculptor group, at about 3.9 Mpc \citep{Karachentsev2003}. The initial survey identified 27 interesting sources in the vicinity of NGC 7793, P13 being the most luminous among them at the time of the observation in 1992, at about $9 \times 10^{38}$ erg s$^{-1}$. Among those 27 ROSAT sources, another source was later identified as a highly variable ULX, NGC 7793 P9 \citep{Hu2018}. Both P9 and P13 lie in the outskirts of the galaxy. P13 has been observed a total of 14 times between 2012 and 2019 with \textit{XMM-Newton}, for a total of over 600ks of exposure. This ULX was shown to harbour a 0.42s pulsar \citep{Furst2016, Israel2017}, and the companion star has been identified as a B9Ia \citep{Motch2014}.

We have discovered another interesting source in this galaxy, NGC 7793 ULX-4. Thanks to the presence of NGC 7793 P13, there have been numerous observations of this galaxy, and consequently of this source. The low number of detections however indicate high variability; taking into account the peak flux and assuming its association with the galaxy implies a highly variable ULX.

\begin{figure}
    \centering
	\includegraphics[width=1\columnwidth]{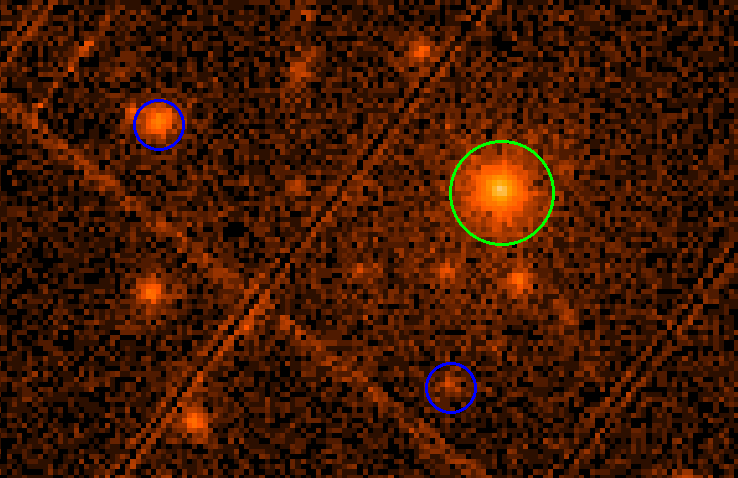}
    \caption{A view from the \textit{XMM-Newton} observation 0693760101. The green circle indicates the newly discovered ULX in its highest detected state. This 40" radius circular region was used to extract the photons in the analysis. The blue circles correspond to the known ULXs in NGC 7793, respectively P9 on the top left and P13 on the bottom right, which was the target of the observation and in a low state at this time. The field of view is about 360"$\times$560", and roughly covers the optical extent of the galaxy.}
    \label{fig:XMMview}
\end{figure}

In this paper, we present the study of this new source, using a wealth of multi-wavelength data from \textit{XMM-Newton}, the Neil Gehrels \textit{Swift} Observatory (hereby \textit{Swift}), and \textit{Chandra} in X-rays and from the \textit{HST} and the VLT Multi Unit Spectroscopic Explorer (hereby MUSE) in the optical. In Section \ref{sec:Data_selection}, we describe the data selection and analysis. The timing and spectral analysis results of the study are then shown in Section \ref{sec:Results}. These results are finally summed up and discussed in Section \ref{sec:Conclusion}.

\section{Data selection \& reduction}\label{sec:Data_selection}

\subsection{XMM Newton}
While looking for long term variability of \textit{XMM-Newton} sources in the \href{http://xmmssc.irap.omp.eu/Catalogue/4XMM-DR9/4XMMv3.pdf}{4XMM-DR9 catalog} \citep{4XMMDR9}, we came upon the extra-nuclear source 4XMM J235747.9-323457 (RA=359.4496, DEC=-32.5826). It was detected in a single \textit{XMM-Newton} observation, ObsID 0693760101 (see Figure~\ref{fig:XMMview}). Since NGC 7793 has been observed by \textit{XMM-Newton} a total of 14 times between May 14$^{\text{th}}$, 2012 and November 22$^{\text{nd}}$, 2019, the HILIGT upper limit server\footnote{\url{http://xmmuls.esac.esa.int/upperlimitserver/}} was employed to determine the flux upper limits in the case of non-detections. We chose a power-law spectral model with a $\Gamma=$ 1.7 slope, a hydrogen column density of $n_H = 3\times10^{-20}$ cm$^{-2}$, and working on the 0.2-12 keV band, the server returned 3$\sigma$ flux upper limits for 9 observations (see Table \ref{tab:observations}). These values were chosen to be similar to the parameters used to determine the fluxes of sources detected in the 4XMM-DR9 catalog. Four observations returned automatically computed detections, which we did not take into account as they were polluted by noise or readout streaks, and consequently are not part of the 4XMM-DR9 catalog. The server also allowed us to retrieve flux upper limits for the \textit{XMM-Newton} Slew survey, on 3 Slew observations (all details are in Table \ref{tab:observations}).

We then studied the single \textit{XMM-Newton} detection of this source, in the observation 0693760101, in May 2012. We used the \textit{XMM Newton} Science Analysis System v.18.0.0\footnote{"Users Guide to the \textit{XMM-Newton} Science Analysis System", Issue 15.0, 2019 (ESA: \textit{XMM-Newton} SOC)} to extract the data from all three EPIC instruments using \texttt{epproc} and \texttt{emproc}, and we applied the following process for the EPIC-pn data (and for the EPIC-MOS data respectively). We removed the high-flaring periods when the events with pattern 0 and with energies greater than 10 keV yielded a rate greater than 0.4 counts.s$^{-1}$ (0.35 counts.s$^{-1}$ for MOS), as is recommended in the \textit{XMM-Newton} data analysis threads. This left about 29 ks from the initial 50 ks. Events were further filtered by removing the bad pixels and the patterns $\leq$4 (resp. patterns $\leq$12). The source spectra were then extracted from a circular region centred on the source with a radius of 40" for all three instruments, in order to reach encircled energy fractions of about 90\%. The background spectra were selected from 60" radius circular regions in empty zones on the same CCD as the source. We generated the redistribution matrices and ancillary files using respectively the \texttt{rmfgen} and \texttt{arfgen} tasks. Finally, all spectra were rebinned using \texttt{grppha}, with bins of at least 20 photons, and the fits were performed using a $\chi^{2}$ statistic.

For the timing analysis, only the EPIC-pn data was used. No  background was subtracted from the data. The data were barycentered using the position of the source and the SAS task \texttt{BARYCEN}.

\subsection{Swift}
A \textit{Swift} counterpart was found in the 2SXPS catalog \citep{Evans2020}, named 2SXPS J235747.8-323458. There are 13 \textit{Swift} detections, allowing us to study the evolution of its luminosity after the outburst. In 181 additional observations, the source was not detected.
The unabsorbed fluxes for each detection were those from the catalog, using a powerlaw with a spectral index of 1.7 and galactic absorption. We used the HILIGT upper limit server to retrieve the flux upper limits for \textit{Swift}, using the same parameters as for \textit{XMM-Newton} but in the 0.3 -- 10.0 keV band. In the same way as for \textit{XMM-Newton}, we removed the server detections that are not part of the 2SXPS catalog. The observations where the source was detected are described in Table \ref{tab:observations}.

For the spectral study, we retrieved the source products from the automatic pipeline \citep{Evans2009}. Our goal was to determine whether there was an evolution of the spectral shape between the high state, as observed by \textit{XMM-Newton}, and the declining state, as observed for some of the \textit{Swift} detections. Since the \textit{Swift} observations cover a large part of the outburst, we stacked the detections into two groups: those corresponding to the high state, for which the catalog gave a luminosity above $10^{39}$ erg.s$^{-1}$; and those corresponding to the fainter state, below $10^{39}$ erg.s$^{-1}$. For both groups, we regrouped the data so each spectral bin contained at least two photons. The fits were performed using C-statistic instead of $\chi^2$, given the low number of photons available.

\subsection{Chandra}
The X-ray observatory \textit{Chandra} \citep{10.1117/12.391545} has observed NGC 7793 four times, but the source went undetected in all occasions. We therefore estimated 3$\sigma$ upper limits on the unabsorbed flux in the 0.5 --7 keV band using the task \texttt{srcflux} in \texttt{CIAO} version 4.12 with CALDB 4.9.3. As for the \textit{Swift} data, we assumed a powerlaw of photon index 1.7 subject to neutral absorption along the line of sight (\textit{n}$_\text{HGal}$ = 3.38 $\times$ 10$^{20}$ cm$^{-2}$). The observation details and the resulting upper limits can be found in Table \ref{tab:observations}.

\subsection{HST}
In order to determine whether our source is indeed located in NGC 7793,  we looked for its optical counterpart. We examined three Hubble Space Telescope observations in the B, V and I bands, which were made on December 10$^\text{th}$, 2003 (details are in Table \ref{tab:observations}). The excellent angular resolution allowed us to look for optical counterparts within the \textit{XMM-Newton} 1$\sigma$ position error zone. The \texttt{astropy} \citep{Astropy2013,Astropy2018} and \texttt{photutils} \citep{Photutils} Python packages were used to compute photometry for the sources within this circle. For each possible counterpart, the source and background fluxes were extracted from circular regions of respective radii 0.08" and 0.16" centred on the counterparts. We assumed a poissonian error on the photon counts. 

\subsection{MUSE}
In addition to the different counterpart candidates we retrieved from the HST data, we took advantage of the fact that the MUSE instrument \citep{Bacon2010} on the {\it Very Large Telescope} (VLT) observed the galaxy NGC 7793 during its Adaptive Optics Science Verification run\footnote{ESO programme 60.A-9188(A), PI Adamo}. We used the cube ADP.2017-06-16T14:34:33.584, which had a 3500s exposure time. Using the Python package \texttt{mpdaf} \citep{MPDAF2016}, we corrected a 3" offset with the HST data, by assuming a correct HST astrometry and using the \texttt{estimate\_coordinate\_offset} function, which looks for the peak of the cross-correlation between both images. We then proceeded to extract the optical spectrum (475--935 nm) of the zone associated with the X-ray source, that contains all the HST counterpart candidates. The lower spatial resolution (PSF$\sim$0.75") precludes us from studying individually each of the sources found in the HST image; the spectrum thus corresponds to the combination of the four candidates. Moreover, this source is quite close to the core of NGC 7793, about 40" away, to be compared with 2' for NGC 7793 P13. Considerable extended emission from the galaxy is thus present in this region, making the background subtraction challenging; as a consequence, the spectrum is very noisy. It was not possible to identify the nature of the stellar companion as was done for instance for NGC 7793 P13 \citep{Motch2014}. It was however possible to check for the presence of strongly redshifted or galactic lines, in order to assess the possibility of a foreground or background optical counterpart.

\begin{figure*}
    \centering
    \includegraphics[width=0.49\textwidth]{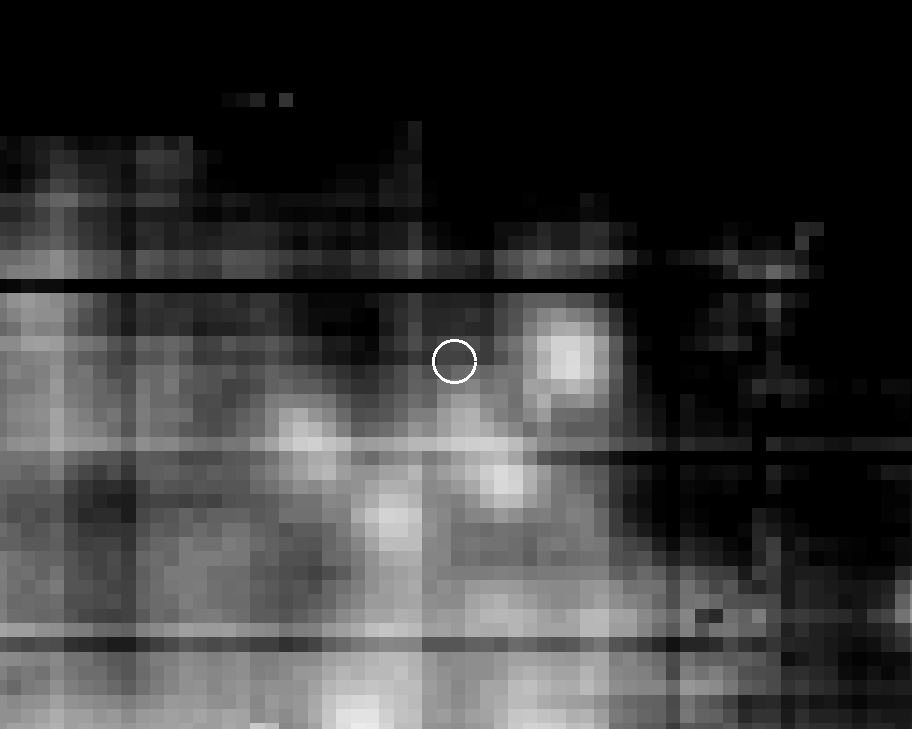}
     \includegraphics[width=0.49\textwidth]{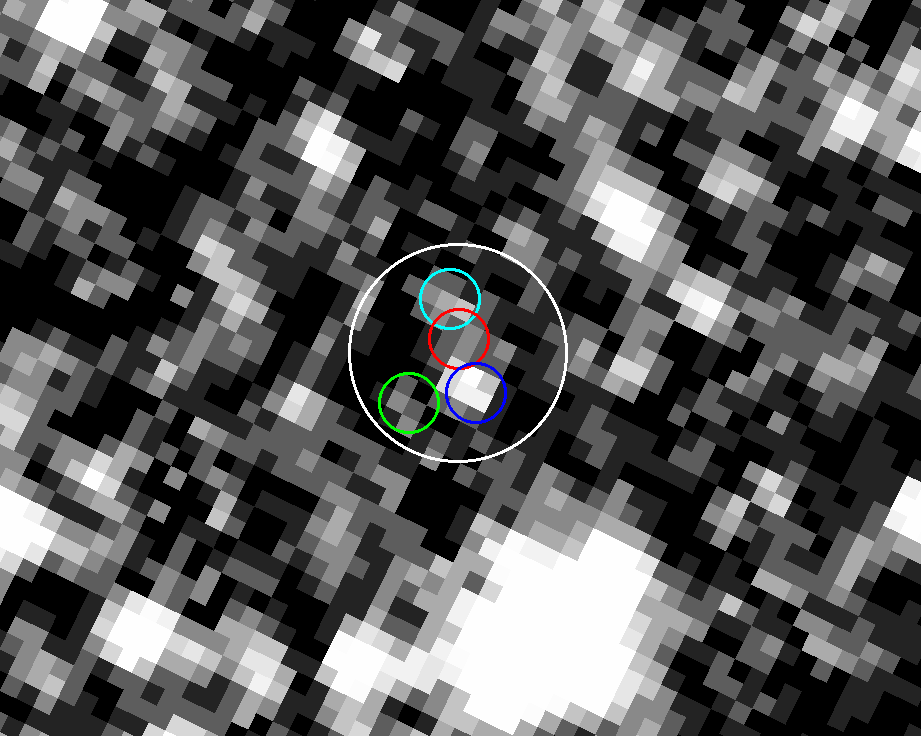}
     \includegraphics[width=0.49\textwidth]{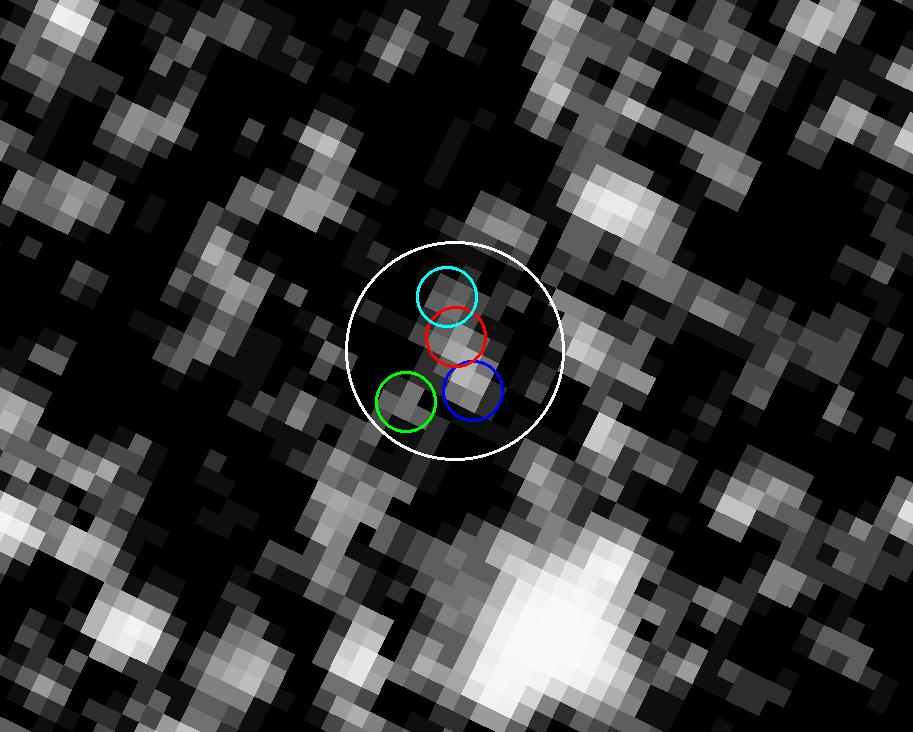}
     \includegraphics[width=0.49\textwidth]{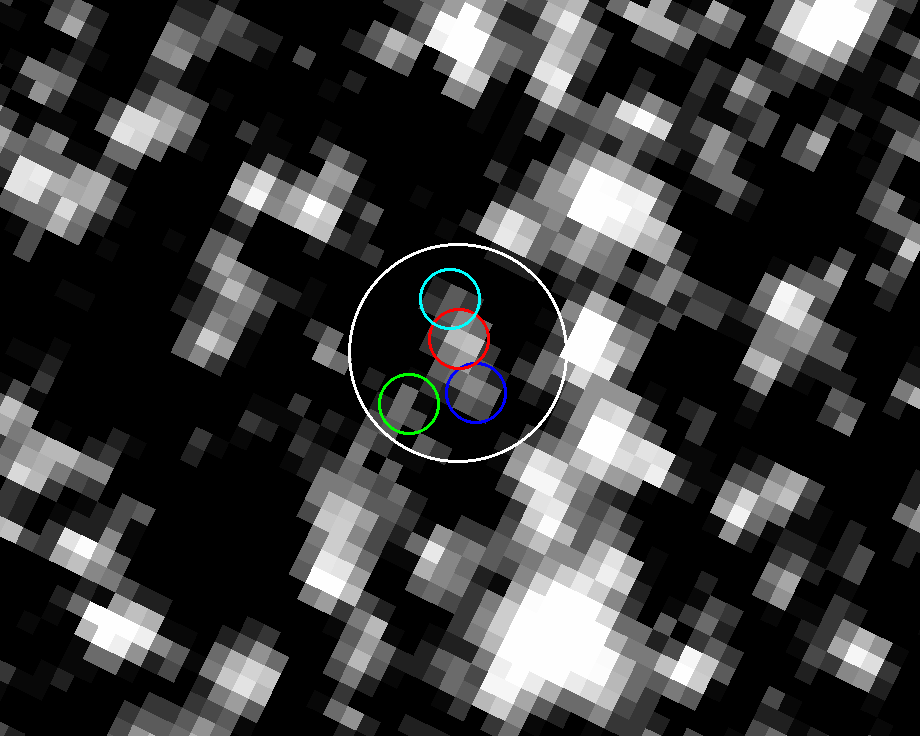}
    \caption{\textit{Top Left}: MUSE data, averaged. The white circle corresponds to the X-ray position with a 1$\sigma$ error radius of 0.29", and the field of view is roughly 10"$\times$13". \textit{Top Right}: HST data in the B band. \textit{Bottom Left}: HST data in the V band \textit{Bottom Right}: HST data in the I band. All the HST images correspond to a field of view of about 2"$\times$2.5", the white circle still corresponds to the X-ray 1$\sigma$ position error. We can see the four possible counterparts, each with a 0.8" circle around it: Blue circle is Source 1; Red circle is Source 2; Cyan circle is Source 3; Green circle is Source 4.}
    \label{fig:Optical}
\end{figure*}

\begin{table*}

\centering

\caption{Log of observations of NGC 7793 ULX-4. For the sake of readability, we did not add the 171 non-detections from \textit{Swift} (among which 10 were considered as detections by the upper limit server but are not part of 2SXPS, and were therefore ignored). The unabsorbed \textit{XMM-Newton} detection flux was computed based on the spectral study we performed. The unabsorbed \textit{Swift} fluxes were taken from 2SXPS, using a fixed 1.7 powerlaw. All \textit{XMM-Newton} and \textit{Swift} upper limits were computed using HILIGT, with a 1.7 powerlaw and $n_H=3\times10^{-20}$ cm$^{-2}$. The upper limits correspond to a $3\sigma$ significance. All \textit{Swift} observations marked with a dagger were used for the spectral study of the low-luminosity phase, as for them the pipeline yielded luminosity below 10$^{39}$ erg.s$^{-1}$. The \textit{Chandra} upper limits were computed using the \texttt{CIAO} tools.}
\begin{tabular}{cccccc}
\hline
\hline
Instrument (Band)& ObsID & Date & Exposure (s) & Detection status & Flux (erg.s$^{-1}$.cm$^{-2}$) \\ \hline \hline
\textit{XMM-Newton} / EPIC pn     &   	0693760101    &  2012-05-14   &  50957 & Detection & $1.89^{+0.09}_{-0.07}\times10^{-12}$ \\
(0.2-12.0 keV)&   	0693760401    &  2013-11-25   &  49000 & Upper Limit & < $9.93\times10^{-15}$\\ 
&   	0748390901    &  2014-12-10   &  50000 & Upper Limit &  < $5.56\times10^{-15}$\\
&   	0781800101    &  2016-05-20   &  53000 & Upper Limit& < $9.18\times10^{-15}$\\
&   	0804670201    &  2017-05-13   &  30400 & Upper Limit& < $1.86\times10^{-14}$\\
&   	0804670301    &  2017-05-20   &  57900 & Upper Limit& < $6.38\times10^{-15}$\\
&   	0804670401    &  2017-05-31   &  34000 & Upper Limit& < $1.24\times10^{-14}$\\
&   	0804670501    &  2017-06-12   &  35100 & Spurious detection&\\ 
&   	0804670601    &  2017-06-20   &  32500 & Spurious detection&\\ 
&   	0804670701    &  2017-11-25   &  53000 & Upper Limit& < $7.47\times10^{-15}$\\
&   	0823410301    &  2018-11-27   &  28000 & Spurious detection&\\ 
&   	0823410401    &  2018-12-27   &  28001 & Upper Limit& < $1.04\times10^{-14}$\\
&   	0840990101    &  2019-05-16   &  46500 & Upper Limit& < $1.44\times10^{-14}$\\
&   	0853981001    &  2019-11-22   &  53500 & Spurious detection&\\ 
\hline
\textit{XMM-Newton} / EPIC pn  - Slew survey   &   	9337600004    &  2018-05-17   &  N/A & Upper Limit & < $2.35\times10^{12}$\\
(0.2-12.0 keV)&   	9348900003    &  2018-12-27   &  N/A & Upper Limit &< $5.10\times10^{-13}$\\
&   	9376400002   &  2020-06-28  &  N/A & Upper Limit& < $6.21\times10^{-13}$\\
\hline
\textit{Swift} / XRT & 00046280001& 2012-04-15 &276&Detection&$(1.15\pm0.55)\times10^{-12}$ \\
(0.2-10.0 keV)& 00046280002$^\dagger$&2012-07-20& 1400&Detection&$(5.38\pm1.64)\times10^{-13}$ \\
& 00046280003&2012-07-26 & 568&Detection&$(1.89\pm0.49)\times10^{-12}$  \\
& 00046280004& 2012-07-30 & 3800&Detection&$(9.84\pm1.15)\times10^{-13}$  \\
& 00091356001& 2012-09-02 &2700&Detection&$(7.94\pm1.34)\times10^{-13}$ \\
& 00091356002& 2012-09-06 &2800&Detection& $(7.96\pm1.35)\times10^{-13}$\\
& 00091356003$^\dagger$&2012-09-10 & 2900&Detection&$(5.10\pm1.06)\times10^{-13}$\\
& 00091356004$^\dagger$&2012-10-15  &3200&Detection&$(5.46\pm1.04)\times10^{-13}$ \\
& 00091356006$^\dagger$&2012-10-21 & 2900&Detection&$(4.11\pm0.98)\times10^{-13}$\\
& 00091356007$^\dagger$&2012-11-10  & 2500&Detection&$(2.36\pm0.87)\times10^{-13}$  \\
& 00046280005$^\dagger$&2012-11-12 & 3300&Detection&$(2.06\pm0.75)\times10^{-13}$ \\
& 00091356008$^\dagger$&2012-11-13  & 2900&Detection&$(2.01\pm0.72)\times10^{-13}$ \\
& 00091356009$^\dagger$&2012-11-21  & 2800&Detection&$(1.38\pm0.63)\times10^{-13}$\\
\hline
\textit{Chandra} / ACIS & 3954 & 2003-09-06 & 48 940 & Upper limit& < $1.58\times10^{-15}$\\
(0.5 -- 7.0 keV)& 14231 & 2011-08-13 & 58 840 & Upper limit&< $1.06\times10^{-15}$\\
& 13439 & 2011-12-25 & 57 770 & Upper limit&< $1.39\times10^{-15}$\\
& 14378 & 2011-12-30 & 24 700 & Upper limit&< $3.39\times10^{-15}$\\\hline
\textit{HST} / ACS-WFC (B band)    &   j8ph0j010   &   2003-12-10   & 680 & N/A & $(4.01\pm0.40 )\times10^{-15}$\\ 
(V band)    &   j8ph0j020   &   2003-12-10   & 680 & N/A &$(2.81\pm0.46 )\times10^{-15}$\\ 
(I band)    &   j8ph0j030   &   2003-12-10   & 430 & N/A &$(1.37\pm0.25 )\times10^{-15}$\\ 
\hline
\textit{ESO} / MUSE (475--935 nm)  & ADP.2017-06-16T14:34:33.584 & 2017-06-16 & 3500 &N/A &N/A \\
\hline
\end{tabular}

\label{tab:observations}
\end{table*}

\section{Results}\label{sec:Results}
\subsection{Long term evolution}\label{sec:LongTerm}
\begin{figure*}
    \centering
    \includegraphics[width=\textwidth]{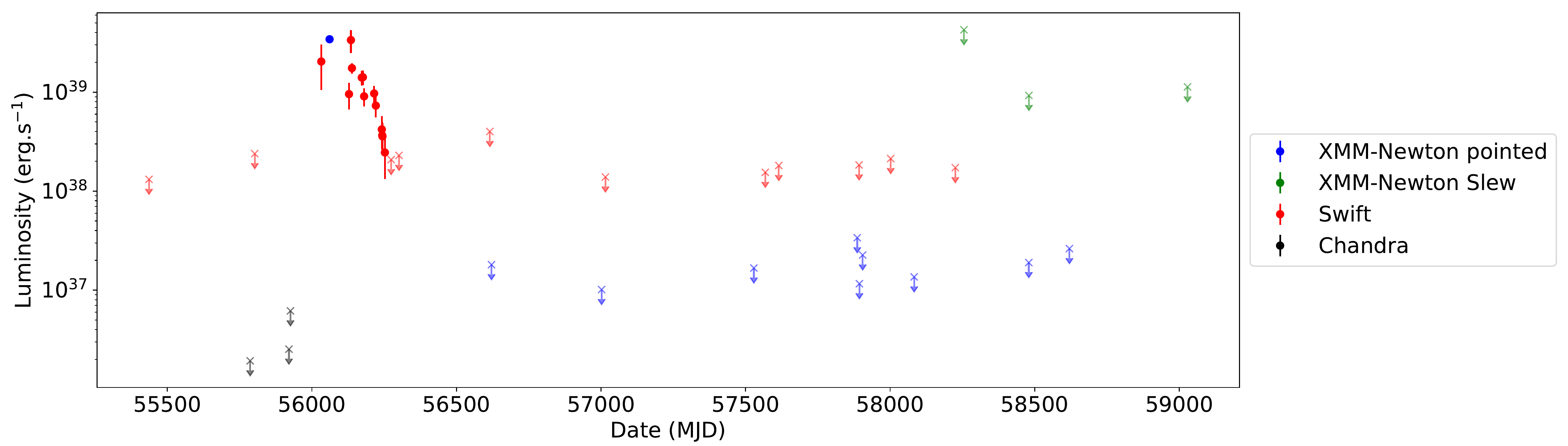}
    \caption{Long-term evolution of the source. The blue points correspond to \textit{XMM-Newton} pointed observations, the red ones to \textit{Swift} data, the black ones to \textit{Chandra}, and the green ones to the \textit{XMM-Newton} slew data. Crosses with arrows correspond to upper limits, while filled dots correspond to detections. The upper limits for \textit{XMM-Newton} (pointed and slew) and \textit{Swift} were computed using the ESA Upper Limit Server. For the sake of readability, out of the 171 \textit{Swift} upper limits we only plotted the lowest ones in time bins corresponding to 300 days (the approximate duration of the peak). The \textit{Chandra} upper limits were computed using \texttt{CIAO}; for the sake of readability, we removed the first \textit{Chandra} observation, 8 years before the peak, with a value similar to the other upper limits. For all data points, the conversion from flux to luminosity was done using a distance to NGC 7793 of 3.9 Mpc \citep{Karachentsev2003}. The lightcurve shows large variability with a high-activity episode in 2012, lasting for about a year, and the absence of any activity since then.}
    \label{fig:Long_term_lightcurve}
\end{figure*}

The long term evolution of the X-ray source is shown in Figure \ref{fig:Long_term_lightcurve}. The combined use of detections and upper limits from different observatories (\textit{XMM-Newton}, \textit{Swift}, \textit{Chandra}) reveals the transient nature of this ULX, on a timescale of about 8 months. The \textit{Chandra} upper limits provide us with a constraint on the rise time of this ULX, with about four months between \textit{Chandra}'s last low-state observation (ObsID 14378 on December 30$^\text{th}$, 2011) and \textit{Swift}'s first high-state observation (DatasetID 00046280001 on April 15$^\text{th}$, 2012). The amplitude of variability is about three orders of magnitude, which is consistent with the most variable ULXs in \cite{Song2020}.

\begin{figure*}
    \centering
    \includegraphics[width=0.48\textwidth]{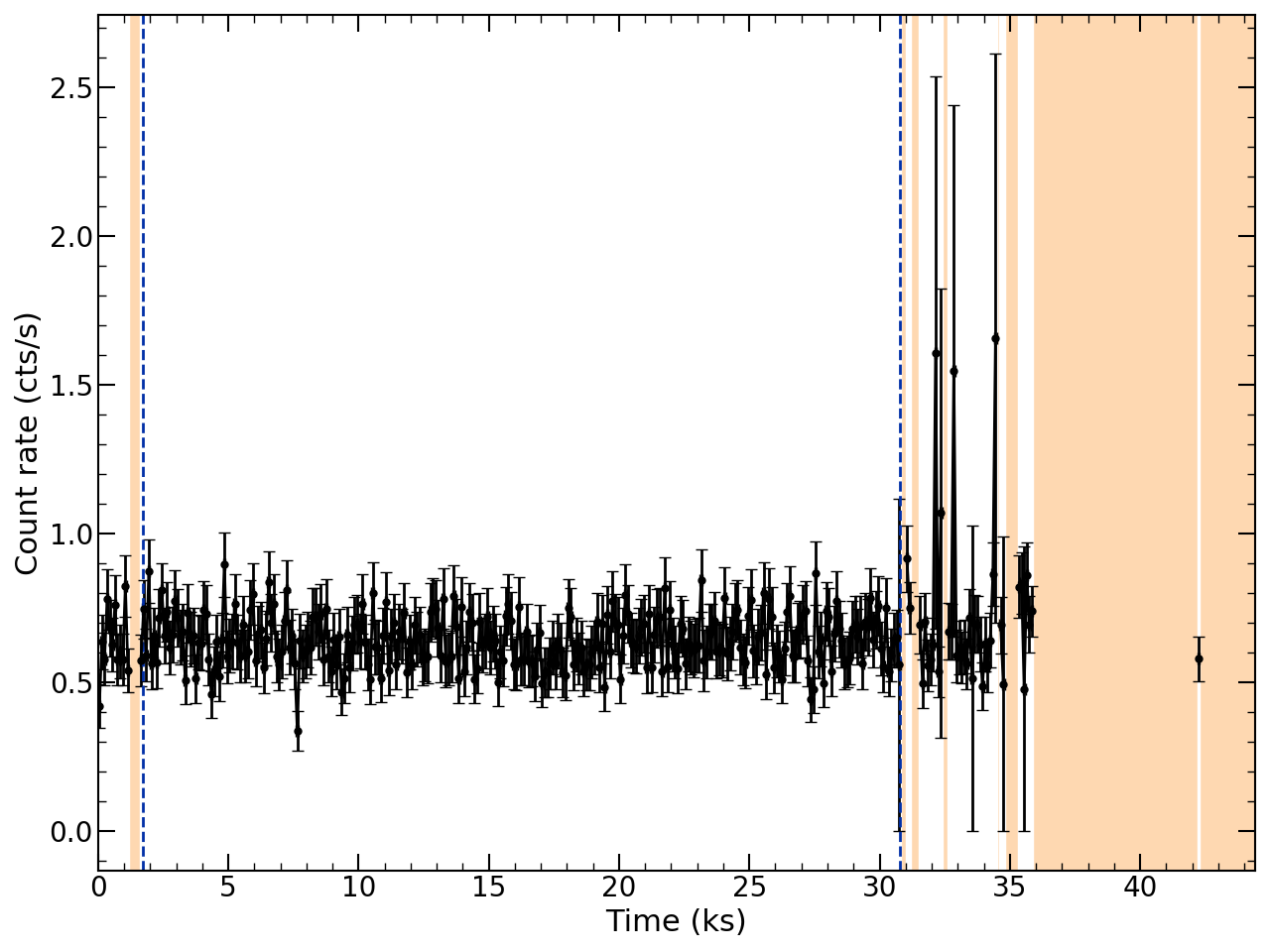}
     \includegraphics[width=0.48\textwidth]{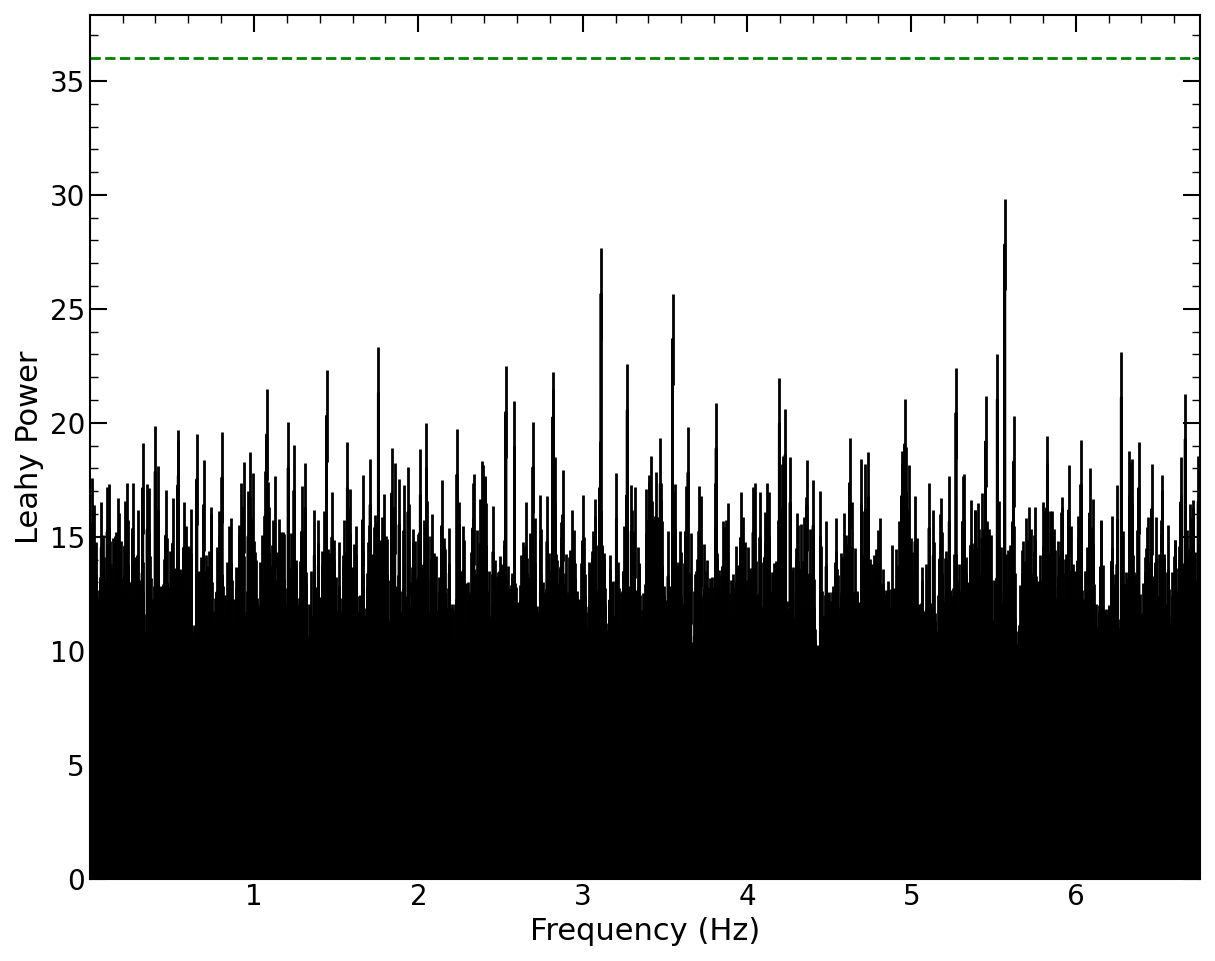}
    \caption{Left: pn lightcurve in the 0.3 -- 10 keV of \thesource\ during the outburst (obsid 0693760101) binned to 100s. 
    The lightcurve has been corrected using \textit{epiclccorr}. Periods filtered by background flaring activity are highlighted in orange (see Section \ref{sec:Data_selection}). The black-dashed lines indicate the data selected for subsequent timing analysis. Right: Power density spectrum from the selected lightcurve segment at the maximal resolution ($\sim$ 74ms). 
    The expected 3$\sigma$ false alarm probability is shown as a green dashed line and is calculated based on Equation 3.7 from \citet{vanderklisFourierTechniquesXray1989}.}
    \label{fig:lightcurve}
\end{figure*}

\subsection{Short term evolution}
In order to analyse the short-term X-ray variability of the source, we focus on the \textit{XMM-Newton} observation 0693760101, where the source was clearly detected. Figure \ref{fig:lightcurve} shows the pn lightcurve corrected using \texttt{epicclcorr} binned to 100s, where small data gaps (t$<$20s) have been replaced by Poisson distributed rates around the mean count rate computed from the GTIs. As can be seen, the source shows little variability throughout the observation. For the subsequent timing analysis, we selected the events in the time window shown in Figure \ref{fig:lightcurve} between the black dashed lines, to avoid artifacts introduced by the large data gaps created by the GTI. This results in a net exposure of $\sim$ 29 ks. 

We next produced two power density spectra (PDS) from the \textit{total} (i.e. source + background) 29 ks uncorrected lightcurve using the Leahy normalization \citep{1983ApJ...266..160L}, as we find that the \texttt{epicclcorr} task can alter the underlying statistics. The first one is produced from the entire lightcurve rebinned at the maximal resolution ($\sim$ 74ms), suitable to look for coherent pulsations up to $\sim$ 7 Hz. For the second one, we bin the lightcurve to 0.5s and average together power spectra of time series of 500s long (see \cite{1988tns..conf...27V} for a review), in order to look for quasi-periodic oscillations similar to those reported by \cite{2019MNRAS.486.2766A}. We find both PDS consistent with being dominated by Poisson noise. The first PDS is shown in Figure \ref{fig:lightcurve} to highlight that no pulsations were identified. 

\subsubsection{Pulse searches}
\label{subsec:PulseSearch}

\begin{figure}
    \centering
    \includegraphics[width=\linewidth]{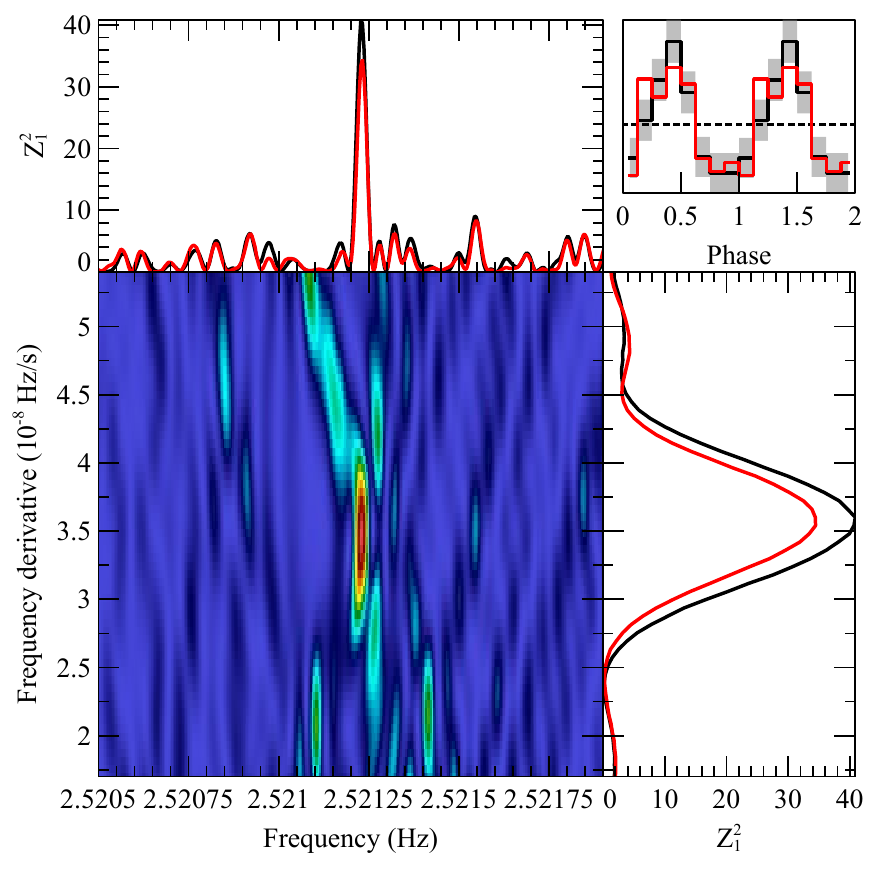}
    \caption{Candidate pulsation at 2.52 Hz. 
    The lines in the upper left panel and in the lower right panel represent the statistics $Z^2_1$ (Rayleigh test) for the best values of frequency and frequency derivative.
    The upper right panel shows the pulse profile with 1-sigma confidence interval in grey.
    Black lines represent the analysis done using the nominal event arrival times; red lines show the same analysis done after summing a random time delta uniformly distributed between $\pm$36.5\,ms (half the frame time)}
    \label{fig:candidate}
\end{figure}

Due to the high spin up rates commonly found in PULXs, more refined timing techniques are generally required to detect any pulsations. We therefore undertook an accelerated pulsation search using the \texttt{HENDRICS} timing software \citep{bachettiHENDRICSHighENergy2018}. We ran an accelerated Fourier-domain search \citep{ransomFourierTechniquesVery2002a}, implemented in the tool \texttt{HENaccelsearch}, between 0.01 and 7 Hz.
To overcome the loss of sensitivity of the power density spectrum (PDS) close to the borders of the spectral bins, we used ``interbinning'', a way to interpolate the signal between two bins using the signal in the adjacent bins \citep{vanderklisFourierTechniquesXray1989}. 
This search technique produces many false positives, due to the alteration of the white noise level of the PDS, that need to be confirmed with independent methods.
We found 88 candidates, with frequencies between 0.01 and 5 Hz.
A few of the ones at the lower boundary of this range turned out to be clear artifacts, and we could reject them by visually inspecting the folded profiles.

We then analyzed the $f-\dot{f}$ parameter space around all candidate frequencies with the $Z^2_1$ (Rayleigh test) and $Z^2_2$ search\footnote{The candidate pulse frequencies being only a few times greater than the frame time, using more than two harmonics would not improve the sensitivity}\citep{buccheriSearchPulsedGammaray1983a}.
To perform this analysis, we used the tool \texttt{HENzsearch} with the \textit{fast} option, that decreases the computational time of the search by a factor of $\sim10$ and automatically optimizes the search in the $\dot{f}$ space by pre-binning the photons in phase and computing the $Z^2_n$ statistic using these bins instead of the individual photons \citep{Huppenkothen_2019}.

As expected due to the use of interbinning, most of the candidates were not confirmed as significant by the $Z^2_n$ search.
However, one candidate stood out, with a nominal significance of $\sim3.4\,\sigma$.
The pulse being consistent with a sinusoid, the candidate is detected best in the $Z^2_1$ test. 
We show the details in Figure~\ref{fig:candidate}.
This candidate has a frequency of 2.5212(1)\,Hz and a frequency derivative of $3.5(2)\cdot10^{-8}$\,Hz/s.
The pulsed fraction is around $12\pm3\%$ over the \textit{XMM-Newton} energy band.

We caution that estimating the significance in such a search is tricky. 
In particular, it is difficult to estimate the correct number of trials after running an accelerated pulsar search where the spectral bins are not completely independent, using interbinning, and then running the $Z^2_2$ search on the most promising candidates.

Another word of caution: the candidate pulsation is only a few times the frame time.
As an additional test, we randomized the event arrival times adding a time delta uniformly distributed between $\pm$36.5\,ms (plus/minus half the frame time of EPIC-pn in full frame mode), and doing so the significance of the detection decreased slightly below $3\sigma$.
Since we are dealing with a small number of counts, it is hard to say whether we artificially diluted the signal of a \textit{bona fide} candidate.

\subsection{Spectral analysis}
\subsubsection{X-ray spectrum}\label{subsec:xrayspec}
\begin{figure}
    \centering
    \includegraphics[width=0.5\textwidth]{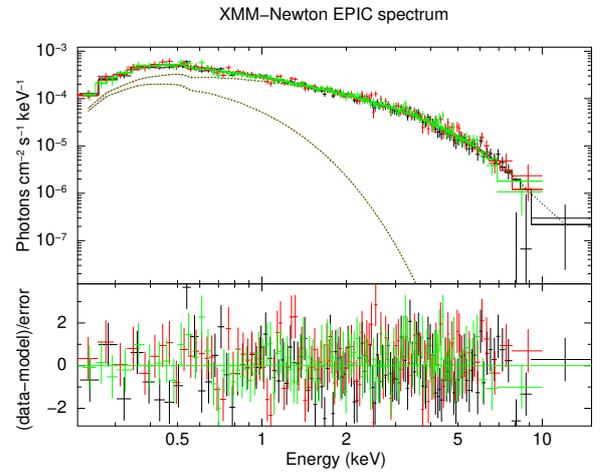}
    \caption{Spectrum obtained from \textit{XMM-Newton} using all the EPIC data: EPIC-pn (black), EPIC-MOS1 (red) and EPIC-MOS2 (green), binned in groups of at least 20 photons, and then simultaneously fitted with an absorbed double multi-colored black body model. For the plot, data was rebinned to 5$\sigma$.}
    \label{fig:Spectrum}
\end{figure}

For the spectral analysis, we used \texttt{xspec} \citep{Arnaud1996}. The spectra obtained from the three \textit{XMM-Newton} instruments were simultaneously fitted in the 0.2--12 keV range with an absorbed double multi-color black body (\texttt{tbabs(diskbb+diskbb)}), which is commonly used for ULXs \citep[see for instance][and references therein]{Koliopanos2017}. The spectrum is shown in Figure \ref{fig:Spectrum}. The results of the fit are shown in Table \ref{tab:FitResults}. The temperature values associated with the two black bodies correspond to typical values found for other ULXs, i.e. a cool (T$\sim$0.2 keV) and a hot (T$\sim$1.5 keV) thermal component \citep[see][for more details]{Koliopanos2017}. To assess the validity of the use of two thermal components, we tried fitting the \textit{XMM-Newton} data with a single absorbed \texttt{diskbb}; this lead to a much larger value of $\chi^2$ (369.61 with 326 degrees of freedom, compared to a 325.32 with 324 degrees of freedom for the double \texttt{diskbb}), with a significant soft excess (residuals>5$\sigma$ below 0.6 keV), thus justifying the use of a second low-energy \texttt{diskbb}. 

The unabsorbed fluxes in different bands were computed using the model component \texttt{cflux}, and freezing the normalisation of any of the black bodies (both yield the same results). The bands used are 0.2 -- 12 keV to assess the total flux, and 0.3 -- 2 keV and 2 -- 10 keV to quantify the hardness of the emission; the hardness ratio being defined as $\big(F_{2-10 keV}-F_{0.3-2 keV}\big)/\big(F_{2-10 keV}+F_{0.3-2 keV}\big)$. These latter bands were chosen in order to be compared with \textit{Swift} data. From the fit, we estimated a 0.2 -- 12.0 keV unabsorbed flux value of $(1.93\pm0.05)\times10^{-12}$ erg.s$^{-1}$.cm$^{-2}$; assuming a distance of 3.9 Mpc \citep{Karachentsev2003}, this implies a peak luminosity of $(3.43^{+0.16}_{-0.12})\times10^{39}$erg.s$^{-1}$.

The \textit{Swift} high state and low state spectra were then compared to the \textit{XMM-Newton} spectrum. For both \textit{Swift} groups, the spectral range used was 0.3 - 7.4 keV, since above this value the source flux dropped below the background. The models were then extrapolated to the complete 0.3 - 10.0 keV range. In both cases, the hydrogen column density was set to the value found in the \textit{XMM-Newton} high state observation, to improve the robustness of the fits on this low-quality data; this value is compatible with the galactic value. 

The \textit{Swift} high state spectrum was fitted with the same absorbed double multi-colored black bodies model as the \textit{XMM-Newton} data, showing comparable spectral parameters to those for the  \textit{XMM-Newton} data.  The large error bars, especially for the low energy black body, are due to the smaller \textit{Swift} collection area and its higher minimum energy.  We estimated the hardness of the spectrum in the same way as for the \textit{XMM-Newton} data, and the results are shown in Table \ref{tab:hardnessEvolution}. 

Then, the low state spectrum was studied. The idea behind this study was to assess the presence of a spectral change along with the change of luminosity. To this effect, we looked for the minimal changes needed in the high state spectrum in order to account for the low state spectrum. We began by freezing the model parameters from the \textit{XMM-Newton} data, and only accounting for the luminosity change by adding a constant offset, which was the only free parameter. This lead to a relatively poor fit, with a C-stat of 68 for 66 degrees of freedom. Then, we removed the constant offset and let both black bodies free, but this lead to poorly constrained parameters, both normalisations being only upper limits; freezing the temperature of the cool black body at 0.1 keV prevented this for the hot black body, but not for the cool one. Finally, seeing the poor constraints on the normalisation of the cool black body, we removed it entirely, allowing to have a good fit statistic with no parameters for which the normalisation could reach zero. This final model still allowed us to evaluate the hardness of the emission, obtained by comparing the fluxes of this model in the 0.3 -- 2 keV and 2 -- 10 keV bands. The spectrum during the decline phase was in the end consistent with a spectral change, and revealed a spectral softening of the source with decreasing luminosity (see Table \ref{tab:hardnessEvolution}).

\begin{figure}
    \centering
    \includegraphics[width=0.5\textwidth]{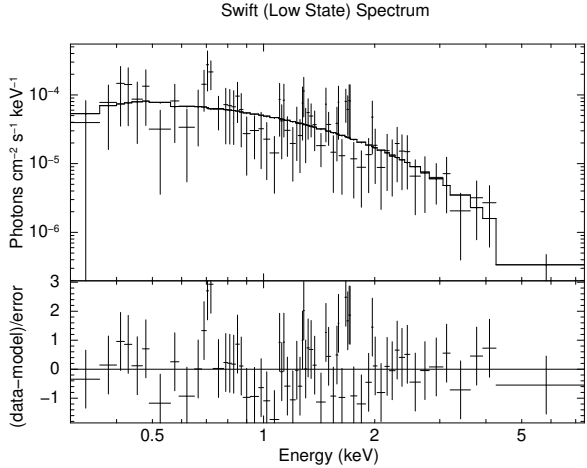}
    \caption{Spectrum obtained from \textit{Swift} low state detections, binned to have at least 2 counts per bin. This spectrum was fitted with a single absorbed multicolor black body using C-statistic. The model was then extrapolated in the full 0.3 - 10.0 keV band.}
    \label{fig:SpectrumSwift}
\end{figure}

\begin{table*}
    \centering
    \caption{Value of the spectral fit parameters. For the high state we used an absorbed double multi-colored black body on both \textit{XMM-Newton} and \textit{Swift} combined observations. For the low state, three different models were implemented on the \textit{Swift} observations, to justify the spectral change: 1) a constant offset fitted from the high state model, 2) a double multi colored black body for which the cool component is fixed at a cold value, and 3) a single multi colored black body. The error bars correspond to 90\% uncertainties.}
    \begin{tabular}{ccccccccc}
    \hline
        State&Spectrum&$n_H$ ($\times10^{20}$cm$^{-2}$) & kT$_1$ (keV)& norm$_1$ & kT$_2$ (keV)& norm$_2 $ ($\times10^{-2}$) & D.o.F & Fit Statistic\\ \hline \hline
        High state&\textit{XMM-Newton} & $4.97^{+1.86}_{-1.49}$ & $0.27^{+0.07}_{-0.04}$ & $2.05^{+3.48}_{-1.38}$ & $1.27^{+0.04}_{-0.04}$ & $3.11^{+0.40}_{-0.40}$ & 324 & $\chi^2$=325.32\\[0.2cm]
        &\textit{Swift}  & $4.97$ (frozen) & $0.19^{+0.22}_{-0.16}$ & $4.13^{+180}_{-3.93}$ & $1.22^{+0.37}_{-0.22}$ & $1.44^{+1.02}_{-0.94}$ & 99 & C-stat = 94.54\\[0.2cm] \hline
        Low state&Constant Offset ($0.14\pm0.01$)  & $4.97$ (frozen) &0.27 (frozen) &2.05 (frozen)  & 1.27 (frozen) &3.11 (frozen)   &66 & C-stat = 68.03\\[0.2cm]
        &Fixed cold \texttt{diskbb}  & $4.97$ (frozen) & 0.1 (frozen) &$13.20_{-13.20}^{+50.02}$  & $0.75^{+0.15}_{-0.12}$&$3.21^{+3.15}_{-1.66}$   &64 & C-stat = 55.02\\[0.2cm]
        &Single \texttt{diskbb} & $4.97$ (frozen) &N/A &N/A  & $0.74^{+0.14}_{-0.11}$&$3.53^{+3.02}_{-1.68}$   &65 & C-stat = 55.27\\
        
        \hline
    \end{tabular}
    
    \label{tab:FitResults}
\end{table*}

\begin{table}
\caption{Flux values in different bands, obtained using the model component \texttt{cflux}, and the corresponding hardness ratios. The \textit{XMM-Newton} data corresponds to the peak, in ObsID 0693760101.}
\resizebox{\columnwidth}{!}{%
\begin{tabular}{cccc}
\hline
Data       & Flux 0.3--2 keV  & Flux 2--10 keV  & HR \\
  & ($\times10^{-13}$ erg.s$^{-1}$) &  ($\times10^{-13}$ erg.s$^{-1}$) &  \\
\hline
\textit{XMM-Newton} &       $8.46^{+0.48}_{-0.67}$            &       $9.48^{+0.25}_{-0.25}$          &  $0.07^{+0.04}_{-0.95}$  \\[0.25cm]
\textit{Swift} (High state) &       $3.58^{+0.53}_{-0.46}$            &       $3.67^{+1.05}_{-0.79}$          &  $0.01^{+0.19}_{-0.19}$  \\[0.25cm]
\textit{Swift} (Low state)      &        $1.19^{+0.20}_{-0.22}$           &        $0.50^{+0.20}_{-0.26}$         &   $-0.33^{+0.22}_{-0.19}$\\
\hline
\end{tabular}
}

\label{tab:hardnessEvolution}
\end{table}

\subsubsection{Optical counterpart}
Four candidate counterparts were found in the HST data as shown in Figure \ref{fig:Optical}. Their absorbed magnitudes are shown in Table \ref{tab:HSTmag}. Sources 1 and 2 have high fluxes in the B band (about $3\times10^{-19}$ erg.s$^{-1}$.cm$^{-2}$), while Source 3 is highest in the I band (about $1.5\times10^{-19}$ erg.s$^{-1}$.cm$^{-2}$); Source 4 has a flatter spectrum. It is impossible at this point to determine precisely which star is the companion to NGC 7793 ULX-4. In order to get a better assessment of the nature of the source, we focused on the X-ray to optical ratio \citep[see][and references therein]{Zolotukhin2016}. For ULXs, it should reach between $10^2$ and $10^3$, about 10 for AGNs and below 0.1 for stars.

Since there are no simultaneous optical and X-ray observation of the source, we first assume a constant optical flux, between the 2003 HST observation and the 2012 ULX outburst; alternatively, we use a \textit{Chandra} non-detection three months before the optical observation to provide an upper limit on the X-ray to optical ratio of the source in its low state. The typical timescale of the outburst (about seven months) is larger  than the interval between the \textit{Chandra} upper limit and the HST observation, which justifies the use of this second method as as quasi-simultaneous multi-wavelength observation. For the optical flux, we considered the December 2003 V band HST observation, computed for each of the optical counterparts.
The resulting approximations of the $F_X/F_\text{opt}$ for the different counterparts and both methods are shown in table \ref{tab:FxFopt}. The approximation of constant optical flux gives very large values of $F_X/F_\text{opt}$, consistent with ULXs and excluding AGNs or stars. Even if we conservatively assumed a simultaneous optical variability of a factor 10$^3$ at the time of the X-ray outburst (which is the order of magnitude of the X-ray variability), the ratio indicates a ULX in the galaxy NGC 7793. The $F_X/F_\text{opt}$ ratio for the low state (using the \textit{Chandra} non-detection) gives an upper limit that is consistent with a ULX in NGC 7793, but does not exclude a background AGN or a foreground star.
\begin{table}
\centering
\caption{Absorbed magnitudes of the four possible HST counterparts. }

\begin{tabular}{ccccc}
\hline
Band & Source 1 & Source 2 & Source 3 & Source 4\\ \hline
B  & $21.24\pm1.19$&$21.86\pm1.54$&$21.86\pm1.54$&$22.41\pm1.92$\\
V & $20.97\pm1.17$&$20.90\pm1.13$& $21.40\pm1.42$&$21.40\pm1.42$\\ 
I   & $20.07\pm1.08$& $19.59\pm0.85$& $20.25\pm1.17$&$20.28\pm1.19$\\
\hline
\end{tabular}
\label{tab:HSTmag}
\end{table}

\begin{table}
\centering
\caption{$F_X/F_\text{opt}$ calculated using two different methods: either assuming a constant optical flux, or using the X-ray flux upper limit quasi-simultaneous to the optical observation (three months difference).}
\begin{tabular}{ccccc}
\hline
$F_X/F_\text{opt}$ method& Source 1  & Source 2&Source 3& Source 4\\ \hline
$F_\text{opt}$ constant& $8.9\times10^6$ &$8.3\times10^6$&$1.3\times10^6$&$1.3\times10^6$\\
$F_X$ upper limit & <$2.1\times10^{3}$&<$2.0\times10^{3}$&<$2.5\times10^{3}$&<$2.5\times10^{3}$\\ 
\hline
\end{tabular}

\label{tab:FxFopt}
\end{table}

An additional step in confirming the association of these sources with the galaxy is the study of their redshift. MUSE's spatial resolution only allows us to study these sources as a whole, preventing us from determining each source's redshift individually. However, the redshift of both $H_\alpha$ and $H_\beta$ lines, determined by fitting gaussian profiles to the spectrum, is compatible with that of NGC 7793, which is $(7.57\pm0.07)\times10^{-3}$ \citep{Lauberts1989}. This further indicates that the source belongs to NGC 7793. The luminosity is therefore compatible with a ULX and the strong variability supports the identification as a transient ULX.

\section{Discussion \& Conclusion}
\label{sec:Conclusion}

The first step in studying this source is confirming its association with NGC 7793 by excluding a background or foreground contamination. Several arguments can be made against a background AGN. Firstly, the amplitude of the X-ray variability, about three orders of magnitude, is unlikely for an AGN \citep[which typically have X-ray variability of the order a factor 10, see][]{McHardy2010}. The X-ray softening during the declining phase, shown in Table~\ref{tab:hardnessEvolution}, is the opposite of what is expected for an AGN, which are typically softer during high-luminosity episodes. Additionally, the peak X-ray to optical ratio is very high for an AGN. Finally, the absence of highly redshifted H$_\alpha$ or H$_\beta$ lines in the MUSE data tends to exclude a background galaxy as well. We then consider a foreground source, either a Galactic X-ray binary or a star. The Galactic latitude of about -77$^{\circ}$ means that if this source is galactic, it could not be further than about 1 kpc from Earth. Taking the peak flux measured by \textit{XMM-Newton}, this yields a peak luminosity below $(2.26\pm0.05)\times10^{32}$ erg.s$^{-1}$. This peak value is too low for X-ray binaries, which are expected to reach about $10^{38}$ erg.s$^{-1}$ \citep{Kuulkers2003}. It would be consistent with a stellar flare, but the duration of the peak (about 300 days) is much longer than what is expected for a star \citep[a few hours at most, ][]{Pye2015}. The peak X-ray to optical ratio is once again too high for a star. All these arguments tend to exclude any foreground or background object. In addition to this, the spectral shape of the peak emission as observed by \textit{XMM-Newton} is consistent with that of a ULX, and the long-term variability is comparable to that of other varying ULXs \citep{Earnshaw2018, Song2020}. This level of variability on such timescales may be due to a central neutron star transitioning from a propeller state \citep{Tsygankov2016}, during which its magnetic field stops the accretion flow and thus limits the luminosity.

The spectral fits allow us to put some constraints on the source; more precisely, the normalisation of the cool \texttt{diskbb} component gives us information about the inner radius of the disk (in km), through the relation $Norm=(R_{\textrm{disk}}/D_{\textrm{10kpc}})^2 \textrm{cos}(i)$, $i$ being the inclination \citep{Mitsuda1984,Makishima1986}. Using this relation for the \textit{XMM-Newton} data, we get that in the high state
$$ R_{\textrm{disk}} = 550\pm320 \textrm{km} / \textrm{cos}(i)^{1/2}.$$
Due to the unknown inclination, this is a lower limit on the inner radius of the disk. Such a value is comparable with that of other ULXs \citep{Koliopanos2017}. Assuming that the inner radius of the disk coincides with the magnetospheric radius $R_{m}$, we can get an estimate of the central neutron star's magnetic field, using the Equation (1) from \cite{Mushtukov2017}:
$$ R_m = 7 \times 10^{7} \Lambda \ m^{1/7} R_{6}^{10/7}B_{12}^{4/7}L_{39}^{-2/7} \text{cm},$$
where $\Lambda$ is a constant depending on the accretion geometry, often taken to be $\Lambda=0.5$ for disk accretion; $m$ is the mass of the neutron star in units of solar masses, taken to be $m=1.4$; $R_{6}=R/10^{6}$ cm is the radius of the neutron star, taken to be $R_{6}=1$; $B_{12}=B/10^{12}$ G is the surface magnetic field strength, and $L_{39}=L/10^{39}$erg\ s$^{-1}$ is the accretion luminosity, taken to be the peak X-ray luminosity of $L_{39}\sim3$. Assuming cos($i$)$\sim$1, this leads to an estimate of the surface magnetic field of the neutron star during the peak at
$$B\sim 3.5^{+4.1}_{-2.5} \times 10^{12} \text{ G} .$$
This value of the magnetic field is once again comparable to that of other ULXs, among them P13, with $B_\text{P13}\sim4.2\times10^{12}$ G \citep{Koliopanos2017}. For the low state, if we again assume that the disk corresponds to the cool black body we can only use the second model from Table \ref{tab:FitResults}, which is the only one with a cool component. Its normalisation, although poorly constrained, is consistent with an increase between the high state and low state; this would mean an increase  in the inner disc radius, as the lower luminosity implies that the accretion rate drops. The ram pressure then diminishes, so the disc can not penetrate as close to the neutron star.

Another point that can be made about this source's declining state spectrum is the possible presence of emission lines. Indeed, taking a look at Figure \ref{fig:SpectrumSwift} reveals that some data bins collectively diverge ($\sim 2\sigma$) from the continuum emission, at $\sim0.7$ keV and $\sim1.6$ keV. Taking into consideration the quality of the data in hand, it is impossible to conclude whether these data spikes are indeed emission lines. However, such emission lines around these energies are expected to be detected in ULXs, generally massively blueshifted \citep{Pinto2017}, thus revealing the presence of a powerful wind at relatively large fractions of $c$.

Once the source was confidently identified as a varying ULX, the rest of our study of NGC 7793 ULX-4 focused on the search for pulsation. As stated above, we find a candidate pulsation at 2.5212(1)\,Hz with an associated frequency derivative of $3.5(2)\cdot10^{-8}$\,Hz/s. Under the hypothesis that this candidate is a real pulsation, the frequency derivative we found would be indicative of orbital motion. 
Such an $\dot{f}$ would indeed be unrealistically high for the accretion torque on a neutron star, given the flux we measured in Section~\ref{subsec:xrayspec}.
According to standard accretion theory \citep[e.g.][]{GL79b,wang_torque_1995}, given a corotation radius $R_{\rm co}$ and a magnetospheric radius $R_{\rm co}$ one expects the frequency derivative to be roughly related to luminosity \citep[see, e.g.][]{Bachetti2020}
$$\dot{f} \propto \dot{L}^{x/7}.$$
where $x$ is 6 for thin disk accretion \`a la \citet{Shakura1973} and 3 for super-Eddington accretion \`a la \citet{kingMassesBeamingEddington2009}.
The PULX with the highest luminosity observed so far, NGC 5907 ULX-1  ($10^{41}$erg s$^-1$, \citealt{Israel20175907}), has $\dot{f}\approx10^{-9}$\,Hz/s.
The peak luminosity of NGC 7793 ULX-4 being much lower, such a high value for the frequency derivative most likely comes from orbital motion. We do not have enough counts to test the presence of a second derivative, which would allow better constraints on the orbital parameters.
The measured $\dot{f}$ is in the range observed for the ULXs with a known orbital period in the range of $\sim$few days \citep{Bachetti2014,rodriguezcastilloDiscoveryPulsarDay2020}.

Like M 82 X-2 \citep{Bachetti2014}, the orbital period of a few days implies an intermediate mass stellar companion for mass transfer to occur. Intermediate mass X-ray binaries have frequently been suggested to be at the origin of ULXs \citep[see e.g.][ and references therein]{misr20}. Indeed, for a neutron star ULX with a luminosity about 10 times Eddington, the most likely system is an intermediate mass X-ray binary with a period of the order of days \citep{misr20}.

The pulse period of NGC 7793 ULX-4 is also surprisingly close to that of P13 \citep[2.4 Hz,][]{Israel2017}, but the $\dot{f}$ values are extremely different ($2.2\times10^{-10}$ Hz.s$^{-1}$ for P13 compared to $3.5\times10^{-8}$ Hz.s$^{-1}$ for ULX-4), which excludes a possible contamination of the observation by P13. Finally, the fact that P13 was in a low-luminosity state at the time of the peak luminosity for ULX-4 (see Figure \ref{fig:XMMview}), and that the two ULXs are separated by 155", further excludes the possibility of a contamination by P13.

Assuming that the pulsation detected for ULX-4 is confirmed, NGC 7793 is the first known galaxy hosting two pulsating ULXs. Another ULX known in this galaxy, P9 \citep{Hu18} has been proposed to contain a black hole based on spectral analysis. However, the strong variability in this source is reminiscent of the large levels of variability observed in PULXs \citep{Song2020}, frequently ascribed to the transition to the propeller regime. A detailed search for pulsations will be the subject of a future paper. Another object exists in  NGC 7793 \citep[P8 in][]{Read1999} that has been proposed to be a third ULX, S26  \citep{sori10}. Whilst it is not currently seen as a ULX using X-ray observations, the power output measured using radio observations of the surrounding environment indicates that this is also a ULX. The radio and X-ray observations support a black hole hypothesis for the accretor. No other sources have been formally identified as ULXs in this galaxy. If there are other ULXs, they are likely to be highly variable, as we have not seen them in decades of observations, again indicative of ULXs containing a neutron star. Such a large proportion of neutron star ULXs stands out from other galaxies which have both persistent and transient ULXs. Why NGC 7793 should be different is unclear. Following population synthesis studies, \cite{wikt19} propose that more neutron star ULXs exist either in 1) galaxies with constant star formation rates, approximately solar metallicities and ages of at least 1 billion years, or 2) in galaxies a long time after a burst of star formation in regions with lower than solar metallicity.  The central regions of NGC 7793 (in which lies ULX-4) fulfill the criteria for the first case \citep{vlaj11} and the outer regions of NGC 7793 (in which lies P13) fulfill the criteria for case two \citep{vlaj11}. This may explain the propensity for neutron star ULXs in NGC 7793.

Other observations of NGC 7793 ULX-4 during an outburst are needed to confirm the pulse period and determine the first and second period derivatives. Further spectral studies over the outburst would help to confirm the physical nature of the processes behind the evolution of luminosity, and would also allow us to rigorously study the presence emission lines and their possible blueshift. Finally, optical observations during the outburst would be useful to identify which of the possible counterparts is the true counterpart. This source has been quiescent since its first and only detected outburst, in May 2012; the constant monitoring of the nearby ULX NGC 7793 P13 will however undoubtedly allow us to detect a potential new high activity period in ULX-4.

\section*{Acknowledgements}

We thank the referee for their helpful comments. EQ and NW thank the CNES for their support. This research made use of Photutils, an Astropy package for detection and photometry of astronomical sources (Bradley et al.
2020).

\section*{Data availability}
The data underlying this article are available in Zenodo, at \url{http://doi.org/10.5281/zenodo.4472907}. The datasets were derived from sources in the public domain: \url{http://nxsa.esac.esa.int/nxsa-web/} for \textit{XMM-Newton}, \url{https://cxc.cfa.harvard.edu/csc/} for \textit{Chandra}, \url{https://www.swift.ac.uk/2SXPS/} for \textit{Swift}, \url{https://mast.stsci.edu/} for HST, and \url{http://archive.eso.org/scienceportal/} for MUSE. All software are in the public domain: \url{https://www.cosmos.esa.int/web/xmm-newton/} for \texttt{SAS},  \url{https://heasarc.gsfc.nasa.gov/docs/software/heasoft/} for \texttt{HEASoft}, \url{https://cxc.cfa.harvard.edu/ciao} for \texttt{CIAO}, and \url{https://hendrics.readthedocs.io/} for \texttt{HENDRICS}. The code used to look for variable sources will soon be made public, as part of a future article.




\bibliographystyle{mnras}
\bibliography{mnras_template} 

\begin{thebibliography}{}
\makeatletter
\relax
\def\mn@urlcharsother{\let\do\@makeother \do\$\do\&\do\#\do\^\do\_\do\%\do\~}
\def\mn@doi{\begingroup\mn@urlcharsother \@ifnextchar [ {\mn@doi@}
  {\mn@doi@[]}}
\def\mn@doi@[#1]#2{\def\@tempa{#1}\ifx\@tempa\@empty \href
  {http://dx.doi.org/#2} {doi:#2}\else \href {http://dx.doi.org/#2} {#1}\fi
  \endgroup}
\def\mn@eprint#1#2{\mn@eprint@#1:#2::\@nil}
\def\mn@eprint@arXiv#1{\href {http://arxiv.org/abs/#1} {{\tt arXiv:#1}}}
\def\mn@eprint@dblp#1{\href {http://dblp.uni-trier.de/rec/bibtex/#1.xml}
  {dblp:#1}}
\def\mn@eprint@#1:#2:#3:#4\@nil{\def\@tempa {#1}\def\@tempb {#2}\def\@tempc
  {#3}\ifx \@tempc \@empty \let \@tempc \@tempb \let \@tempb \@tempa \fi \ifx
  \@tempb \@empty \def\@tempb {arXiv}\fi \@ifundefined
  {mn@eprint@\@tempb}{\@tempb:\@tempc}{\expandafter \expandafter \csname
  mn@eprint@\@tempb\endcsname \expandafter{\@tempc}}}

\bibitem[\protect\citeauthoryear{{Arnaud}}{{Arnaud}}{1996}]{Arnaud1996}
{Arnaud} K.~A.,  1996, in {Jacoby} G.~H.,  {Barnes} J.,  eds,  Astronomical
  Society of the Pacific Conference Series Vol. 101, Astronomical Data Analysis
  Software and Systems V. p.~17

\bibitem[\protect\citeauthoryear{{Astropy Collaboration} et~al.,}{{Astropy
  Collaboration} et~al.}{2013}]{Astropy2013}
{Astropy Collaboration} et~al., 2013, \mn@doi [\aap]
  {10.1051/0004-6361/201322068}, \href
  {https://ui.adsabs.harvard.edu/abs/2013A&A...558A..33A} {558, A33}

\bibitem[\protect\citeauthoryear{{Astropy Collaboration} et~al.,}{{Astropy
  Collaboration} et~al.}{2018}]{Astropy2018}
{Astropy Collaboration} et~al., 2018, \mn@doi [\aj] {10.3847/1538-3881/aabc4f},
  \href {https://ui.adsabs.harvard.edu/abs/2018AJ....156..123A} {156, 123}

\bibitem[\protect\citeauthoryear{{Atapin}, {Fabrika}  \&
  {Caballero-Garc{\'\i}a}}{{Atapin} et~al.}{2019}]{2019MNRAS.486.2766A}
{Atapin} K.,  {Fabrika} S.,   {Caballero-Garc{\'\i}a} M.~D.,  2019, \mn@doi
  [MNRAS] {10.1093/mnras/stz1027}, \href
  {https://ui.adsabs.harvard.edu/abs/2019MNRAS.486.2766A} {486, 2766}

\bibitem[\protect\citeauthoryear{Bachetti}{Bachetti}{2018}]{bachettiHENDRICSHighENergy2018}
Bachetti M.,  2018, Astrophysics Source Code Library, p. ascl:1805.019

\bibitem[\protect\citeauthoryear{{Bachetti} et~al.,}{{Bachetti}
  et~al.}{2013}]{Bachetti2013}
{Bachetti} M.,  et~al., 2013, \mn@doi [\apj] {10.1088/0004-637X/778/2/163},
  \href {https://ui.adsabs.harvard.edu/abs/2013ApJ...778..163B} {778, 163}

\bibitem[\protect\citeauthoryear{{Bachetti} et~al.,}{{Bachetti}
  et~al.}{2014}]{Bachetti2014}
{Bachetti} M.,  et~al., 2014, \mn@doi [Nat] {10.1038/nature13791}, \href
  {https://ui.adsabs.harvard.edu/abs/2014Natur.514..202B} {514, 202}

\bibitem[\protect\citeauthoryear{{Bachetti} et~al.,}{{Bachetti}
  et~al.}{2020}]{Bachetti2020}
{Bachetti} M.,  et~al., 2020, \mn@doi [\apj] {10.3847/1538-4357/ab6d00}, \href
  {https://ui.adsabs.harvard.edu/abs/2020ApJ...891...44B} {891, 44}

\bibitem[\protect\citeauthoryear{{Bacon} et~al.,}{{Bacon}
  et~al.}{2010}]{Bacon2010}
{Bacon} R.,  et~al., 2010, in Ground-based and Airborne Instrumentation for
  Astronomy III. p. 773508, \mn@doi{10.1117/12.856027}

\bibitem[\protect\citeauthoryear{{Bacon}, {Piqueras}, {Conseil}, {Richard}  \&
  {Shepherd}}{{Bacon} et~al.}{2016}]{MPDAF2016}
{Bacon} R.,  {Piqueras} L.,  {Conseil} S.,  {Richard} J.,   {Shepherd} M.,
  2016, {MPDAF: MUSE Python Data Analysis Framework} (\mn@eprint {ascl}
  {1611.003})

\bibitem[\protect\citeauthoryear{Bradley et~al.,}{Bradley
  et~al.}{2020}]{Photutils}
Bradley L.,  et~al., 2020, astropy/photutils: 1.0.0,
  \mn@doi{10.5281/zenodo.4044744}, \url
  {https://doi.org/10.5281/zenodo.4044744}

\bibitem[\protect\citeauthoryear{Buccheri et~al.,}{Buccheri
  et~al.}{1983}]{buccheriSearchPulsedGammaray1983a}
Buccheri R.,  et~al., 1983, A\&A, 128, 245

\bibitem[\protect\citeauthoryear{{Carpano}, {Haberl}, {Maitra}  \&
  {Vasilopoulos}}{{Carpano} et~al.}{2018}]{Carpano2018}
{Carpano} S.,  {Haberl} F.,  {Maitra} C.,   {Vasilopoulos} G.,  2018, \mn@doi
  [\mnras] {10.1093/mnrasl/sly030}, \href
  {https://ui.adsabs.harvard.edu/abs/2018MNRAS.476L..45C} {476, L45}

\bibitem[\protect\citeauthoryear{{Colbert} \& {Mushotzky}}{{Colbert} \&
  {Mushotzky}}{1999}]{Colbert1999}
{Colbert} E. J.~M.,  {Mushotzky} R.~F.,  1999, \mn@doi [\apj] {10.1086/307356},
  \href {https://ui.adsabs.harvard.edu/abs/1999ApJ...519...89C} {519, 89}

\bibitem[\protect\citeauthoryear{{Earnshaw}, {Roberts}  \&
  {Sathyaprakash}}{{Earnshaw} et~al.}{2018}]{Earnshaw2018}
{Earnshaw} H.~P.,  {Roberts} T.~P.,   {Sathyaprakash} R.,  2018, \mn@doi
  [MNRAS] {10.1093/mnras/sty501}, \href
  {https://ui.adsabs.harvard.edu/abs/2018MNRAS.476.4272E} {476, 4272}

\bibitem[\protect\citeauthoryear{{Evans} et~al.,}{{Evans}
  et~al.}{2009}]{Evans2009}
{Evans} P.~A.,  et~al., 2009, \mn@doi [MNRAS]
  {10.1111/j.1365-2966.2009.14913.x}, \href
  {https://ui.adsabs.harvard.edu/abs/2009MNRAS.397.1177E} {397, 1177}

\bibitem[\protect\citeauthoryear{{Evans} et~al.,}{{Evans}
  et~al.}{2020}]{Evans2020}
{Evans} P.~A.,  et~al., 2020, \mn@doi [APJs] {10.3847/1538-4365/ab7db9}, \href
  {https://ui.adsabs.harvard.edu/abs/2020ApJS..247...54E} {247, 54}

\bibitem[\protect\citeauthoryear{{F{\"u}rst} et~al.,}{{F{\"u}rst}
  et~al.}{2016}]{Furst2016}
{F{\"u}rst} F.,  et~al., 2016, \mn@doi [APJl] {10.3847/2041-8205/831/2/L14},
  \href {https://ui.adsabs.harvard.edu/abs/2016ApJ...831L..14F} {831, L14}

\bibitem[\protect\citeauthoryear{Ghosh \& Lamb}{Ghosh \& Lamb}{1979}]{GL79b}
Ghosh P.,  Lamb F.~K.,  1979, ApJ, 234, 296

\bibitem[\protect\citeauthoryear{{Hu}, {Kong}, {Ng}  \& {Li}}{{Hu}
  et~al.}{2018a}]{Hu2018}
{Hu} C.-P.,  {Kong} A. K.~H.,  {Ng} C.~Y.,   {Li} K.~L.,  2018a, \mn@doi [APJ]
  {10.3847/1538-4357/aad5e2}, \href
  {https://ui.adsabs.harvard.edu/abs/2018ApJ...864...64H} {864, 64}

\bibitem[\protect\citeauthoryear{Hu, Kong, Ng  \& Li}{Hu et~al.}{2018b}]{Hu18}
Hu C.-P.,  Kong A. K.~H.,  Ng C.-Y.,   Li K.~L.,  2018b, \mn@doi [The
  Astrophysical Journal] {10.3847/1538-4357/aad5e2}, 864, 64

\bibitem[\protect\citeauthoryear{Huppenkothen et~al.,}{Huppenkothen
  et~al.}{2019}]{Huppenkothen_2019}
Huppenkothen D.,  et~al., 2019, \mn@doi [The Astrophysical Journal]
  {10.3847/1538-4357/ab258d}, 881, 39

\bibitem[\protect\citeauthoryear{Israel et~al.,}{Israel
  et~al.}{2017a}]{Israel20175907}
Israel G.~L.,  et~al., 2017a, \mn@doi [Science] {10.1126/science.aai8635}, 355,
  817

\bibitem[\protect\citeauthoryear{{Israel} et~al.,}{{Israel}
  et~al.}{2017b}]{Israel2017}
{Israel} G.~L.,  et~al., 2017b, \mn@doi [MNRAS] {10.1093/mnrasl/slw218}, \href
  {https://ui.adsabs.harvard.edu/abs/2017MNRAS.466L..48I} {466, L48}

\bibitem[\protect\citeauthoryear{Kaaret, Feng  \& Roberts}{Kaaret
  et~al.}{2017}]{Kaaret2017}
Kaaret P.,  Feng H.,   Roberts T.~P.,  2017, \mn@doi [Annual Review of
  Astronomy and Astrophysics] {10.1146/annurev-astro-091916-055259}, 55, 303

\bibitem[\protect\citeauthoryear{{Karachentsev} et~al.,}{{Karachentsev}
  et~al.}{2003}]{Karachentsev2003}
{Karachentsev} I.~D.,  et~al., 2003, \mn@doi [A\&A]
  {10.1051/0004-6361:20030170}, \href
  {https://ui.adsabs.harvard.edu/abs/2003A&A...404...93K} {404, 93}

\bibitem[\protect\citeauthoryear{King}{King}{2009}]{kingMassesBeamingEddington2009}
King A.~R.,  2009, \mn@doi [MNRAS Let.] {10.1111/j.1745-3933.2008.00594.x},
  393, L41

\bibitem[\protect\citeauthoryear{{Koliopanos}, {Vasilopoulos}, {Godet},
  {Bachetti}, {Webb}  \& {Barret}}{{Koliopanos} et~al.}{2017}]{Koliopanos2017}
{Koliopanos} F.,  {Vasilopoulos} G.,  {Godet} O.,  {Bachetti} M.,  {Webb}
  N.~A.,   {Barret} D.,  2017, \mn@doi [\aap] {10.1051/0004-6361/201730922},
  \href {https://ui.adsabs.harvard.edu/abs/2017A&A...608A..47K} {608, A47}

\bibitem[\protect\citeauthoryear{{Kuulkers}, {den Hartog}, {in't Zand},
  {Verbunt}, {Harris}  \& {Cocchi}}{{Kuulkers} et~al.}{2003}]{Kuulkers2003}
{Kuulkers} E.,  {den Hartog} P.~R.,  {in't Zand} J.~J.~M.,  {Verbunt} F.~W.~M.,
   {Harris} W.~E.,   {Cocchi} M.,  2003, \mn@doi [\aap]
  {10.1051/0004-6361:20021781}, \href
  {https://ui.adsabs.harvard.edu/abs/2003A&A...399..663K} {399, 663}

\bibitem[\protect\citeauthoryear{{Lauberts} \& {Valentijn}}{{Lauberts} \&
  {Valentijn}}{1989}]{Lauberts1989}
{Lauberts} A.,  {Valentijn} E.~A.,  1989, {The surface photometry catalogue of
  the ESO-Uppsala galaxies}

\bibitem[\protect\citeauthoryear{{Leahy}, {Darbro}, {Elsner}, {Weisskopf},
  {Sutherland}, {Kahn}  \& {Grindlay}}{{Leahy}
  et~al.}{1983}]{1983ApJ...266..160L}
{Leahy} D.~A.,  {Darbro} W.,  {Elsner} R.~F.,  {Weisskopf} M.~C.,  {Sutherland}
  P.~G.,  {Kahn} S.,   {Grindlay} J.~E.,  1983, \mn@doi [APJ] {10.1086/160766},
  \href {https://ui.adsabs.harvard.edu/abs/1983ApJ...266..160L} {266, 160}

\bibitem[\protect\citeauthoryear{{Long}, {Helfand}  \& {Grabelsky}}{{Long}
  et~al.}{1981}]{Long1981}
{Long} K.~S.,  {Helfand} D.~J.,   {Grabelsky} D.~A.,  1981, \mn@doi [APJ]
  {10.1086/159222}, \href
  {https://ui.adsabs.harvard.edu/abs/1981ApJ...248..925L} {248, 925}

\bibitem[\protect\citeauthoryear{{Makishima}, {Maejima}, {Mitsuda}, {Bradt},
  {Remillard}, {Tuohy}, {Hoshi}  \& {Nakagawa}}{{Makishima}
  et~al.}{1986}]{Makishima1986}
{Makishima} K.,  {Maejima} Y.,  {Mitsuda} K.,  {Bradt} H.~V.,  {Remillard}
  R.~A.,  {Tuohy} I.~R.,  {Hoshi} R.,   {Nakagawa} M.,  1986, \mn@doi [\apj]
  {10.1086/164534}, \href
  {https://ui.adsabs.harvard.edu/abs/1986ApJ...308..635M} {308, 635}

\bibitem[\protect\citeauthoryear{{Makishima} et~al.,}{{Makishima}
  et~al.}{2000}]{Makishima2000}
{Makishima} K.,  et~al., 2000, \mn@doi [APJ] {10.1086/308868}, \href
  {https://ui.adsabs.harvard.edu/abs/2000ApJ...535..632M} {535, 632}

\bibitem[\protect\citeauthoryear{{McHardy}}{{McHardy}}{2010}]{McHardy2010}
{McHardy} I.,  2010, {X-Ray Variability of AGN and Relationship to Galactic
  Black Hole Binary Systems}.
p.~203, \mn@doi{10.1007/978-3-540-76937-8_8}

\bibitem[\protect\citeauthoryear{{Misra}, {Fragos}, {Tauris}, {Zapartas}  \&
  {Aguilera-Dena}}{{Misra} et~al.}{2020}]{misr20}
{Misra} D.,  {Fragos} T.,  {Tauris} T.~M.,  {Zapartas} E.,   {Aguilera-Dena}
  D.~R.,  2020, \mn@doi [\aap] {10.1051/0004-6361/202038070}, \href
  {https://ui.adsabs.harvard.edu/abs/2020A&A...642A.174M} {642, A174}

\bibitem[\protect\citeauthoryear{{Mitsuda} et~al.,}{{Mitsuda}
  et~al.}{1984}]{Mitsuda1984}
{Mitsuda} K.,  et~al., 1984, \pasj, \href
  {https://ui.adsabs.harvard.edu/abs/1984PASJ...36..741M} {36, 741}

\bibitem[\protect\citeauthoryear{{Motch}, {Pakull}, {Soria}, {Gris{\'e}}  \&
  {Pietrzy{\'n}ski}}{{Motch} et~al.}{2014}]{Motch2014}
{Motch} C.,  {Pakull} M.~W.,  {Soria} R.,  {Gris{\'e}} F.,   {Pietrzy{\'n}ski}
  G.,  2014, \mn@doi [Nat] {10.1038/nature13730}, \href
  {https://ui.adsabs.harvard.edu/abs/2014Natur.514..198M} {514, 198}

\bibitem[\protect\citeauthoryear{{Mushtukov}, {Suleimanov}, {Tsygankov}  \&
  {Ingram}}{{Mushtukov} et~al.}{2017}]{Mushtukov2017}
{Mushtukov} A.~A.,  {Suleimanov} V.~F.,  {Tsygankov} S.~S.,   {Ingram} A.,
  2017, \mn@doi [\mnras] {10.1093/mnras/stx141}, \href
  {https://ui.adsabs.harvard.edu/abs/2017MNRAS.467.1202M} {467, 1202}

\bibitem[\protect\citeauthoryear{{Pinto} et~al.,}{{Pinto}
  et~al.}{2017}]{Pinto2017}
{Pinto} C.,  et~al., 2017, \mn@doi [\mnras] {10.1093/mnras/stx641}, \href
  {https://ui.adsabs.harvard.edu/abs/2017MNRAS.468.2865P} {468, 2865}

\bibitem[\protect\citeauthoryear{{Pye}, {Rosen}, {Fyfe}  \&
  {Schr{\"o}der}}{{Pye} et~al.}{2015}]{Pye2015}
{Pye} J.~P.,  {Rosen} S.,  {Fyfe} D.,   {Schr{\"o}der} A.~C.,  2015, \mn@doi
  [\aap] {10.1051/0004-6361/201526217}, \href
  {https://ui.adsabs.harvard.edu/abs/2015A&A...581A..28P} {581, A28}

\bibitem[\protect\citeauthoryear{Ransom, Eikenberry  \& Middleditch}{Ransom
  et~al.}{2002}]{ransomFourierTechniquesVery2002a}
Ransom S.~M.,  Eikenberry S.~S.,   Middleditch J.,  2002, \mn@doi [The
  Astronomical Journal] {10.1086/342285}, 124, 1788Figure

\bibitem[\protect\citeauthoryear{{Read} \& {Pietsch}}{{Read} \&
  {Pietsch}}{1999}]{Read1999}
{Read} A.~M.,  {Pietsch} W.,  1999, A\&A, \href
  {https://ui.adsabs.harvard.edu/abs/1999A&A...341....8R} {341, 8}

\bibitem[\protect\citeauthoryear{Rodr{\'i}guez~Castillo
  et~al.,}{Rodr{\'i}guez~Castillo
  et~al.}{2020}]{rodriguezcastilloDiscoveryPulsarDay2020}
Rodr{\'i}guez~Castillo G.~A.,  et~al., 2020, \mn@doi [The Astrophysical
  Journal] {10.3847/1538-4357/ab8a44}, 895, 60

\bibitem[\protect\citeauthoryear{{Sathyaprakash} et~al.,}{{Sathyaprakash}
  et~al.}{2019}]{Sathyaprakash2019}
{Sathyaprakash} R.,  et~al., 2019, \mn@doi [\mnras] {10.1093/mnrasl/slz086},
  \href {https://ui.adsabs.harvard.edu/abs/2019MNRAS.488L..35S} {488, L35}

\bibitem[\protect\citeauthoryear{{Shakura} \& {Sunyaev}}{{Shakura} \&
  {Sunyaev}}{1973}]{Shakura1973}
{Shakura} N.~I.,  {Sunyaev} R.~A.,  1973, A\&A, \href
  {https://ui.adsabs.harvard.edu/abs/1973A&A....24..337S} {500, 33}

\bibitem[\protect\citeauthoryear{{Song}, {Walton}, {Lansbury}, {Evans},
  {Fabian}, {Earnshaw}  \& {Roberts}}{{Song} et~al.}{2020}]{Song2020}
{Song} X.,  {Walton} D.~J.,  {Lansbury} G.~B.,  {Evans} P.~A.,  {Fabian} A.~C.,
   {Earnshaw} H.,   {Roberts} T.~P.,  2020, \mn@doi [MNRAS]
  {10.1093/mnras/stz3036}, \href
  {https://ui.adsabs.harvard.edu/abs/2020MNRAS.491.1260S} {491, 1260}

\bibitem[\protect\citeauthoryear{{Soria}, {Pakull}, {Broderick}, {Corbel}  \&
  {Motch}}{{Soria} et~al.}{2010}]{sori10}
{Soria} R.,  {Pakull} M.~W.,  {Broderick} J.~W.,  {Corbel} S.,   {Motch} C.,
  2010, \mn@doi [\mnras] {10.1111/j.1365-2966.2010.17360.x}, \href
  {https://ui.adsabs.harvard.edu/abs/2010MNRAS.409..541S} {409, 541}

\bibitem[\protect\citeauthoryear{{Tsygankov}, {Mushtukov}, {Suleimanov}  \&
  {Poutanen}}{{Tsygankov} et~al.}{2016}]{Tsygankov2016}
{Tsygankov} S.~S.,  {Mushtukov} A.~A.,  {Suleimanov} V.~F.,   {Poutanen} J.,
  2016, \mn@doi [MNRAS] {10.1093/mnras/stw046}, \href
  {https://ui.adsabs.harvard.edu/abs/2016MNRAS.457.1101T} {457, 1101}

\bibitem[\protect\citeauthoryear{{Vlaji{\'c}}, {Bland-Hawthorn}  \&
  {Freeman}}{{Vlaji{\'c}} et~al.}{2011}]{vlaj11}
{Vlaji{\'c}} M.,  {Bland-Hawthorn} J.,   {Freeman} K.~C.,  2011, \mn@doi [\apj]
  {10.1088/0004-637X/732/1/7}, \href
  {https://ui.adsabs.harvard.edu/abs/2011ApJ...732....7V} {732, 7}

\bibitem[\protect\citeauthoryear{{Walton} et~al.,}{{Walton}
  et~al.}{2013}]{Walton2013}
{Walton} D.~J.,  et~al., 2013, \mn@doi [\apj] {10.1088/0004-637X/779/2/148},
  \href {https://ui.adsabs.harvard.edu/abs/2013ApJ...779..148W} {779, 148}

\bibitem[\protect\citeauthoryear{{Walton} et~al.,}{{Walton}
  et~al.}{2015}]{Walton2015}
{Walton} D.~J.,  et~al., 2015, \mn@doi [APJ] {10.1088/0004-637X/799/2/122},
  \href {https://ui.adsabs.harvard.edu/abs/2015ApJ...799..122W} {799, 122}

\bibitem[\protect\citeauthoryear{Wang}{Wang}{1995}]{wang_torque_1995}
Wang Y.-M.,  1995, \mn@doi [ApJL] {10.1086/309649}, 449, L153

\bibitem[\protect\citeauthoryear{{Webb} et~al.,}{{Webb} et~al.}{2020}]{4XMMDR9}
{Webb} N.~A.,  et~al., 2020, \mn@doi [\aap] {10.1051/0004-6361/201937353},
  \href {https://ui.adsabs.harvard.edu/abs/2020A&A...641A.136W} {641, A136}

\bibitem[\protect\citeauthoryear{{Weisskopf}, {Tananbaum}, {Van Speybroeck}  \&
  {O'Dell}}{{Weisskopf} et~al.}{2000}]{10.1117/12.391545}
{Weisskopf} M.~C.,  {Tananbaum} H.~D.,  {Van Speybroeck} L.~P.,   {O'Dell}
  S.~L.,  2000, in {Truemper} J.~E.,  {Aschenbach} B.,  eds,  Society of
  Photo-Optical Instrumentation Engineers (SPIE) Conference Series Vol. 4012,
  X-Ray Optics, Instruments, and Missions III. pp 2--16 (\mn@eprint {arXiv}
  {astro-ph/0004127}), \mn@doi{10.1117/12.391545}

\bibitem[\protect\citeauthoryear{Wiktorowicz, Lasota, Middleton  \&
  Belczynski}{Wiktorowicz et~al.}{2019}]{wikt19}
Wiktorowicz G.,  Lasota J.-P.,  Middleton M.,   Belczynski K.,  2019, \mn@doi
  [The Astrophysical Journal] {10.3847/1538-4357/ab0f27}, 875, 53

\bibitem[\protect\citeauthoryear{{Wilson-Hodge} et~al.,}{{Wilson-Hodge}
  et~al.}{2018}]{Wilson-Hodge2018}
{Wilson-Hodge} C.~A.,  et~al., 2018, \mn@doi [\apj] {10.3847/1538-4357/aace60},
  \href {https://ui.adsabs.harvard.edu/abs/2018ApJ...863....9W} {863, 9}

\bibitem[\protect\citeauthoryear{{Zolotukhin}, {Webb}, {Godet}, {Bachetti}  \&
  {Barret}}{{Zolotukhin} et~al.}{2016}]{Zolotukhin2016}
{Zolotukhin} I.,  {Webb} N.~A.,  {Godet} O.,  {Bachetti} M.,   {Barret} D.,
  2016, \mn@doi [\apj] {10.3847/0004-637X/817/2/88}, \href
  {https://ui.adsabs.harvard.edu/abs/2016ApJ...817...88Z} {817, 88}

\bibitem[\protect\citeauthoryear{{van der Klis}}{{van der
  Klis}}{1988}]{1988tns..conf...27V}
{van der Klis} M.,  1988, in Timing Neutron Stars. pp 27--70

\bibitem[\protect\citeauthoryear{{van der Klis}}{{van der
  Klis}}{1989}]{vanderklisFourierTechniquesXray1989}
{van der Klis} M.,  1989, in Timing {{Neutron Stars}}: Proceedings of the
  {{NATO Advanced Study Institute}} on {{Timing Neutron Stars}} Held {{April}}
  4-15. p.~27

\makeatother
\end{thebibliography}



\newpage





\bsp	
\label{lastpage}
\end{document}